\documentclass[12pt,oneside,english]{article}

\usepackage{epsfig}

\usepackage[T1]{fontenc}
\usepackage[latin1]{inputenc}
\usepackage{babel}
\usepackage{setspace}

\makeatletter

\begin{document}
\centerline{\large Computer simulations discussed in physical terms and
terminology }

\bigskip \bigskip

 \centerline{\large   D. Bar$^a$}
 
 \bigskip \bigskip
 
 {$a$  \bf Department of Physics, Bar Ilan University, Ramat Gan,
Israel}

\bigskip

\begin{abstract}   \noindent 
{\it    
 As known,  any numerical   simulation  is composed of two parts: (1) the initial 
part of writing the relevant code and (2)  the running of this code 
on the computer screen. The second part of running the program is extensively
discussed theoretically and technically in the relevant literature. 
In this work   we pay special attention to the less discussed first 
part  and show that it may be discussed in a terminology and notation which
describe physical phenomena. As examples we discuss the  two cases of simulating
 (1) the harmonic
oscillator and (2) the electron-photon interaction which  results in the known Lamb
shift.   }     
\end{abstract}
\noindent
{\bf \underline{Keywords}}: Computer Simulation, Stochastic Quantization,
Harmonic Oscillator, 

\hspace{1.60 cm}  Lamb Shift

\bigskip \noindent 
\protect \section{Introduction \label{sec1}}
\noindent
The problem of validating scientific theories through numerical simulations 
have been discussed from several points of view 
\cite{Feynman1,Naylor,Kleindorfer}. 
There is now almost no scientist (physicist, chemist, biologist etc) 
that 
does not use the powerful means of numerical simulations as a 
necessary tool in
his research.  Moreover, it is 
 accepted \cite{Naylor,Kleindorfer} that if some scientific theory  
is found in
 its numerical simulation version  to be valid on the 
screen then generally it is valid also outside it.
 Thus, a correspondence  may be  drawn between the various stages  
in physical
 theories of 
first proposing the scientific theory (writing the relevant equations), 
and then testing it through experiments   
 to the analogous 
steps in computer simulations of first writing the  program and  then running  
  it on the computer screen. 
 This correspondence  between the two processes is especially emphasized in the 
 experimentation  stage  except for the  differences  
 due to their different characters. That is, whereas 
  the  physical  theories are proved or refuted  through real 
experiments performed in the (three-dimensional)  laboratory, 
 the 
relevant "experiments" in the numerical simulations  are  the
running of the involved programs  on  
the (two-dimensional) 
computer screen.   
 \par
First of all  we note that  a programmer who wants to numerically  simulate any 
process    
 may accomplish his task by writing any of a large number of different codes 
  which   are all equivalent for obtaining the same simulation. 
   Also, when he begin to write his program 
   there is no way to predict beforehand the final code  (from all the possible
   ones) so the writing
   stage of any numerical simulation has some unpredictability 
   related to it.   \par   In this 
   work we are, especially, interested in   
the  programming procedure and in the possible mathematical description of it. 
For this purpose we use the known fact that  the  written code    
  of the numerical simulation of any process must contain the full and detailed 
 description of the simulated phenomena. That is, 
  only when one  introduces   into the code  in advance all  the  details  
 of the simulated system that he may expect an appropriate simulation of it 
   when  this program is later run on the computer screen.  
   Thus,  since the programmer must be very careful  in correctly
   describing in his code the behaviour of the simulated system 
    one may suppose that the code-writing itself 
    evolutes   in such a manner that it reflects  
    the real evolution     of the simulated process.      That is, referring 
to the correspondence between
numerical simulation and physical theories, we assume that as the latter
describe 
 real  processes  as rates of change of some
variables in the infinitesimal (or infinite) limits of other dependent variables
so  the simulation of this process may be
described in an analogous way.   \par
 One may analyze the numerical simulations from the 
    point of view of two interacting systems;  the simulated system and 
    the one that enables, through writing the relevant code, the numerical 
    simulation of the former. 
    The dominant and active system in this supposed interaction is of course 
    the simulated one which actually dictate the content of the written 
    code which should, as remarked, represent and  describe the 
    simulated system. Thus, using the terminology of Synergetics (see, for 
    example, page 195 in \cite{Haken1}), the programming act is "enslaved" by 
    the simulated system and may be actually discussed in terms of it. 
    This is the basic and central property  of Synergetics which have 
    been successfully applied not only to Physics bu also to Chemistry, 
    Biology and other exact disciplines. That is, one may describe the 
    interaction between any two heavily interacting systems from the 
    exclusive standpoint of the dominant system. In other words, the 
    evolution and development of the 
   "enslaved"  \cite{Haken1} and passive process of writing the code may be 
   entirely discussed 
   in terms of the 
   "slaving" 
    and dominant simulated system as done here. \par 
  Thus,   for 
  taking into account   
 the fact that there are  many possible different codes 
   which fulfill the same  
 numerical
 tasks we  introduce, as done in other analogous situations which involve  
 different  possible ways to obtain the same result     
 (see, for example, \cite{Lax,Horwitz}),  an extra variable.    
 We use here  the known Parisi-Wu-Namiki Stochastic
Quantization (SQ) method  \cite{Parisi,Namiki}  by which an additional 
 variable  have  been introduced  into either the
Langevin equation \cite{Coffey} or the Fokker-Plank one 
\cite{Risken}. In this  formalism one     
assumes that some (generally unknown) stochastic process 
\cite{Kannan} 
occurs in the extra
dimension of the additional variable and  the equilibrium physical
situations are approached  in the limit of the elimination of this variable   
which is obtained   by equating all its different values  to each
other and
taking to infinity \cite{Parisi,Namiki}.   This formalism is 
appropriate for the
program-writing stage of the simulation which may evolute along different
equivalent routes and, therefore,  
  the unpredictable element  of it corresponds to the mentioned
  stochastic process.    
 The numerical equilibrium stage is obtained when the  writing process
 ends and one remains with the finished program. 
  \par
  For a suitable analysis 
 of the possibilities allowed at the code-writing stage we discuss  a large 
ensemble of     
programmers which   all try to perform the same numerical simulation.  
     We calculte the 
probability to find all or most of them obtaining   not only the same  
final simulation  but also writing the same code which truly represents the 
simulated phenomena.   
That is, we want to find  the path integral \cite{Feynman2} correlation 
and  the conditions which must
 be fulfilled in order to obtain a large probability for finding  them with the 
 same numerical evolution which really describes the simulated system. 
 Thus,  our ultimate test will be to find out if at 
   the equilibrium stage after they have finished writing their codes 
   the obtained simulation correctly represents the simulated real systems.  
    This indeed will be  shown in the following sections 
by the obtained expression for the correlation between the ensemble of codes 
which turns out to be very similar to the correspnding expression for an
ensemble of the real simulated systems.   \par       
In Section 2 we present the SQ method and relate it to numerical simulation.  
In Section 3 we restrict  the
 discussion to the  computer simulations which simulate  the physical harmonic
oscillator.     We show, by discussing, as remarked, the simulation process in
terms of the simulated harmonic oscillator, 
that  the expression obtained for the correlation between the
computers of the ensemble are very similar to 
the known expression for the correlation between the
members of an ensemble of {\it real}  quantum harmonic oscillators.  \par 
In Section 4 we discuss the  numerical    simulation of 
 the known Lamb shift 
 process \cite{Haken,Mahan,Lamb} and   show, using 
the
SQ formalism, the Fokker-Plank equation \cite{Risken}, and the classical 
     Feynman diagram   
\cite{Namiki,Roepstorff,Masao} 
   that  at the  numerical equilibrium state  one may obtain for the 
   correlation, 
as for  the harmonic oscillator case,  
     the analogous known expression obtained in the framework  of 
     quantum field
theory  \cite{Mahan,Mattuck}.    In 
Section  5 we  further analyze and summarize  the discussion.    
    \bigskip 
  \pagestyle{myheadings}
\markright{ REPRESENTATION  OF THE  SIMULATION PROCESS ....}
  
\protect \section{Representation of the simulation process as a stochastic
Langevin equation \label{sec1}}
As remarked, the code of any simulation process must be written so as to include
and describe 
 all the details of the simulated system if one wants the simulation to
truly represents it.  Thus,  one may expect that 
 the programming act evolutes and  
 developes  in a manner which reflects  the corresponding evolution 
  of the
simulated systems.  And   since, as noted,   these real 
   systems  are  mathematically described by taking rates of 
   changes so the numerical simulation of them may likewise be discussed in such
   a manner.  One  have only  to take into account the 
   additional functionality,
   taken care of by the extra variable (denoted $s$),   that by running different 
   codes one 
   may obtain the same numerical result. Thus, if we denote the process of 
   writing the code by  the programmer $i$  by $q_i$, where $ 1 \le i \le N$, 
      then   these $q_i$ 
   may be 
    analyzed  by taking the rates of
   their changes with respect to $s$, that is, as the generalized
Langevin equation \cite{Coffey}   
\begin{equation} \label{e1} 
\frac{\partial q_i(s,t,x)}{\partial s}=K_i(q (s,t,x))+\eta_i (s,t,x),  \ \ \ \ \
\ \ i=1, 2, \ldots N, 
\end{equation}
where $N$ denotes the remarked $N$-member ensemble of programmers which all try
to perform the same numerical simulation. 
The  
  $\eta_i$'s  denote
stochatic processes in the  variable $s$. As remarked, these 
processes 
stand  here for the described  unpredictable nature  of the programming act 
where one 
can not predict beforehand  the final code. 
      The variables $q_i$   depends    
 upon  $s$ and upon the spatial-time axes $(x,t)$,  where  $x$ denotes the two
 dimensional spatial axes of the screen since these numerical simulations occur
 on the screen and $t$ is the time. 
 The
 $K_i$ are given in the SQ theory by 
    \cite{Parisi,Namiki} \begin{equation} \label{e2}
K_i(q(s,t,x)=-(\frac{\partial S_i[q]}{\partial q})_{q=q(s,t,x)},   \end{equation}
where $S_i$ are  the actions $S_i=\int dsL_i(q,\dot q)$ that determine the 
forms of $q_i$  and $L_i$ are their Lagrangians. 
In order to discuss the ``evolution'' of the code-writing  process 
  which  determines the simulation  on the screen,  we
consider the  time and $s$ intervals $(t_0,t)$,  $(s_0,s)$ and divide
each of them into $N$ subintervals $(t_0,t_1)$, $(t_1,t_2)$, \ldots
$(t_{N-1},t)$
  and   $(s_0,s_1)$, $(s_1,t_2)$, \ldots $(s_{N-1},s)$. 
  The subdivision of paths into small subintervals is basic and central 
    in the path integral method not only in the quantum version of it but 
    also in its application to classical systems. 
    This subdivision is very  important  
    here  for the two discussed examples of the harmonic oscillator in Section 2
     and 
    the Lamb shift in Section 4. Since only by applying and calculating the relevant 
    expressions (the Langevin equation here and the Fokker-Plank equation in the
    next section) in these 
    subintervals the  results shown  
    in the following are obtained.
  We assume that the
Langevin Eq (\ref{e1}) is satisfied for each member of the ensemble at each 
subinterval with the following Gaussian constraints \cite{Namiki}
 \begin{equation} \label{e3} <\!\eta_i(t_r,s_r)\!>=0, \ \ \ <\!\eta_i(t_r,s_r)\eta_j(\grave t_r,\grave s_r)\!>=
2\alpha \delta_{ij}\delta (t_r-\grave t_r)\delta (s_r-\grave s_r), \end{equation} 
where the angular brackets denote an ensemble average with the Gaussian distribution. 
The $r$ signifies the $N$ subintervals of each member and the $i$, $j$ denote 
these members where $N \geq i \ne j \geq 1$. Since, as remarked, the written
code  should  fully describe the evolution of the specific
simulated process  the $\alpha$  have different
meanings which depend upon the identity of this process and the context in which 
Eqs (\ref{e1}) and (\ref{e3}) 
are used. Thus, in the classical regime $\alpha$ is \cite{Namiki} 
$\alpha=\frac{k_{\beta}T}{f}, $where $k_{\beta}$, $T$, and $f$ are respectively
the Boltzman constant, the temperature in Kelvin units and the relevant friction
force. In the quantum regime $\alpha$ is identified \cite{Namiki} with the
Plank constant $\hbar$.   
  We note that  using Eq (\ref{e1})  together with the specific
constraints from Eq (\ref{e3}) enables one \cite{Namiki} to  discuss 
 a large number of different classical 
and quantum phenomena. 
  It has been shown  
\cite{Namiki} that  
  the right hand side 
  of  Eq (\ref{e3}) may be
written as   
\begin{equation} \label{e4} P_{\eta_i}(y)dy=
\prod_i\frac{1}{\sqrt{2\pi(<\!\eta_i\!>)^2}}
\exp(-\frac{y_i^2}{2(<\!\eta_i\!>)^2})dy_i, \end{equation} 
which is the probability to have a value of $\eta_i$ in $(y_i,y_i+dy)$
\cite{Namiki}, where 
\begin{equation} \label{e5} y_i=\frac{\partial q_i(s,t,x)}{\partial
s}-K_i(q_i(s,t,x)) \end{equation}
We want to calculate  the Green's functions $\Delta_{ij\ldots}(t_1,t_2,\ldots)$ 
which  determine the correlation among the members of the ensemble 
 \cite{Namiki}.   Thus, one may define, as in \cite{Namiki}, the Green's 
functions which depend also upon the variable $s$. 
 \begin{eqnarray} && \Delta_{ij\ldots}(t_0,s_0,t_1,s_1,\ldots)=
 <\!q_i(t_0,s_0)q_j(t_1,s_1)
 \ldots\!>= \label{e6} \\ && = C\int Dq(t,s)q_i(t_0,s_0)q_j(t_1,s_1)
 \ldots \exp(-\frac{S_i(q(t,s))}{\alpha}),  \nonumber
 \end{eqnarray} 
 where $S_i$ are  the actions $S_i=\int ds L_i(q,\dot q)$, $C$ is a
 normalization constant,  and 
$Dq(t,s)=\prod_{i=1}^{i=n}dq_i(t,s)$. As seen from the last equation the 
$\Delta_{ij\ldots}(t_1,t_2,\ldots)$ were expressed as path integrals
\cite{Feynman2}. 
 Note that the quantum Feynman measure $e^{\frac{iS(q)}{\hbar}}$ is replaced in
 Eq (\ref{e6}) and in the following Eq (\ref{e7})  
 by $e^{-\frac{S(q)}{\alpha}}$ as required for the classical path integrals
 \cite{Namiki,Roepstorff,Swanson}.   \par
  It can be seen that when the $s$'s are different for    the members of the ensemble
so that each have its specific $S_i(q(s_i,t))$, $K_i(q(s_i,t))$, 
and $\eta_i(s_i,t)$
 the correlation in (\ref{e6}) is obviously zero. 
Thus, in order to have a nonzero value for the probability to find a large part
of the ensemble writing the same numerical code   we have to consider
the stationary configuration where, as remarked, all the $s$ values are equated
to each other and eliminated. For that  matter we take  account of the fact 
that the dependence upon $s$ and $t$ is through $q$ so  this ensures
\cite{Namiki} that this dependence is expressed through the $s$ and $t$
differences. For example, referring to the members $i$ and $j$ the correlation
between them is $\Delta_{ij}(t_i-t_j,s_i-s_j)$, so that for eliminating the $s$
variable from the correlation function one equates all these different $s$'s to
each other.   We, thus,  obtain the following stationary equilibrium correlation 
\cite{Namiki} 
\begin{eqnarray}  && \Delta_{ij\ldots}(t_0,s_0,\ldots)_{st}=
 <\!q_i(t_0,s_0)q_j(t_1,s_1)
 \ldots\!>_{st}=C\int Dq(t)q_i(t_0)q_j(t_1) \ldots \label{e7} \\   
 && \ldots \exp(-\frac{S(q)}{\alpha}),  \nonumber 
 \end{eqnarray}
 where the subscript of $st$  denotes the stationary configuration.  
In other words, the equilibrium   correlation in our case is obtained 
when all the different $s$ values which  give rise to different  
 programs 
and so to different simulations  are equated to each other in which case one remains
with the  set of programs which simulate the same system. 
 \par
 Thus, if all  the    members of the
ensemble  write  the same   version of the program which is mathematically
described as having   
  similar
actions $S$ (in which the $s$ values, which denote here different codes,  
 are equated to each other)  one finds 
with a large probability these  members,   
 in the later equilibrium stage,   with    the
same  code.     That is,  introducing the same similar actions into the
corresponding path integrals one finds this mentioned large probability. 
This is  
exemplified  in the following Section 3  for the harmonic oscillator and in Section 4 
for the energy shift case. \par 
\bigskip

 \bigskip 
  \pagestyle{myheadings}
\markright{ THE NUMERICAL SIMULATIONS OF THE HARMONIC OSCILLATOR}
       
\noindent

\protect \section{The numerical simulations of the harmonic oscillator  \label{sec3}}
As an application of the former discussion 
 we calculate the correlation of the $N$ member ensemble with respect to
the numerical simulations of the harmonic oscillator. That is, we calculate
the  probability to find   all or most of the programmers writing 
 the same  code of   the
harmonic oscillator which truly represents the harmonic oscillator.  
 At the initial time $t_0$ when all the
 programmers just begin to write their respective versions of the harmonic
 oscillator code  they are certainly identical to each other.  
  That is,  all the programmers begin from the same common starting 
  point (denoted $q^0$)
   and    
  we want to
find the probability that  they end    at the later time $t$  with 
the same  code  (denoted $q^{(2N-1)}$) which  refers  to the  
harmonic oscillator.  We assume, for convenient mathematical representation of
 the following discussion,  a $2N$ member ensemble and also a subdivision of each 
 of the intervals $(t_0,t)$ and  $(s_0,s)$ into $2N$ subintervals   
  $(t_0,t_1), (t_1,t_2), 
\ldots (t_{2N-1},t)$ and   
$(s_0,s_1), (s_1,s_2), \ldots (s_{2N-1},s)$.  We  write the  Langevin 
equation (\ref{e1}) 
 for the subintervals $(t_{k-1},t_k)$ and $(s_{k-1},s_k)$ in the form  
 \cite{Namiki}
\begin{equation} q_i^k-q_i^{k-1}-K_i(q^{k-1})(s_k-s_{k-1})=d\eta_i^{k-1} 
\label{e8}
\end{equation} 
 The appropriate $K_i$ for the harmonic oscillator is 
\cite{Namiki}  
\begin{equation} \label{e9} K_i(q^{k-1}(t_k,s_k))=m\frac{\partial}{\partial t}
 (\frac{dq_i}{dt_k})-(\frac{\partial V(q_i^k)}{\partial
q})_{q=q(t,s)},  \end{equation}
where $\frac{dq_i}{dt_k}\approx \frac{q_i^k-q_i^{k-1}}{t_k-t_{k-1}}$. 
The  $d\eta_i(s)$ are conditioned  as \cite{Namiki}  $$<\!d\eta_i(s)\!>=0,\ \ \ 
<\!d\eta_i(s)d\eta_j(\grave s)\!>=\left\{ \begin{array}{ll} 0 & {\rm for~} s \ne \grave s \\
2\alpha\delta_{ij}ds & {\rm for~} s=\grave s  \end{array} \right. $$ 
where the $\alpha$ is as discussed after Eq (\ref{e3}) and the  probability from Eq 
(\ref{e4}) assumes the following form for the harmonic oscillator 
\cite{Namiki}  \begin{eqnarray}  && 
P(q^k,t_k,s_k|q^{k-1},t_{k-1},s_{k-1})=
(\frac{1}{\sqrt{2\pi(2\alpha(s_k-s_{k-1}))}})^{2N} 
\cdot \label{e10} \\ && \cdot \exp(-\sum_i\frac{(q_i^k-q_i^{k-1}-K_i(q^{k-1})(s_k-s_{k-1}))^2}
{2(2\alpha (s_k-s_{k-1}))}), \nonumber 
\end{eqnarray}
which is the probability that the $d\eta_i^{k-1}$ from the right hand side of 
Eq (\ref{e8}) take the 
values at its left hand side \cite{Namiki} and the index $i$ runs over 
the $2N$ members of the
ensemble. Note that Eqs (\ref{e8})-(\ref{e10}) are discussed in \cite{Namiki}
without relating the variable $s$ to any numerical simulation.  
 Here, we
relate the variable $s$ to the possible numerical codes of the harmonic 
oscillator which
are written, as remarked, so as to reflect and describe all the details of it. 
Thus,  the  evolution of  the code-writing process should reflect   that  
of the simulated harmonic oscillator according to Eqs (\ref{e8})-(\ref{e10}). 
This is the meaning of saying that the right hand side of Eq (\ref{e8}), which
represents the  unpredictability of the code-writing process, should truly
reflects the left hand side of it which represents the harmonic oscillator.   
A Markov process \cite{Kannan}    in which $\eta(s)$ does not
correlate with its history is always assumed for these correlations. Eq
(\ref{e10}) is the probability that the ensemble is found  at  
$t_{k}$ and $s_k$ with 
the harmonic oscillator code  $q^{k}$ if at  $t_{k-1}$ and  $s_{k-1}$ it was at 
the similar 
   $q^{k-1}$. Note that by  the word code we do  not necessarily mean 
the complete program of the harmonic oscillator, that is, we use this word even 
for  a very small part
of it.  The probability for the entire
interval that the ensemble is found at  $t$ and $s$ to be 
with the same harmonic
oscillator  code $q^{(2N-1)}$ if at the initial  $t_0$,  $s_0$ it 
was at  $q^0$ is  \cite{Namiki} \begin{eqnarray}  
&& P(q^{(2N-1)},t_N,s_N|q^{0},t_0,s_{0})=\int \cdots \int \cdots 
\int P(q^{(2N-1)},t_N,s_N|q^{(2N-2)},t_{(2N-2)},s_{(2N-2)}) \cdots \nonumber  \\ && \cdots 
P(q^k,t_k,s_k|q^{k-1},t_{k-1},s_{k-1}) 
 \cdots P(q^1,t_1,s_1|q^{0},t_0,s_{0})dq^{(2N-1)}\cdots dq^k \cdots dq^1 
 \label{e11}   \end{eqnarray}   
In order to be able to solve the integrals in the former equation we first 
substitute
from Eq (\ref{e9}) into Eq (\ref{e8}). Thus,  dividing the result by the 
infinitesimal interval
$s_k-s_{k-1}=\delta s$, writing for $V(q)$ the quantum mechanical potential
energy $V(q_i)=\frac{1}{2}mw_0q_i^2$ with the eigenvalues
$E_{\nu}=w_0(\nu+\frac{1}{2})$ $\nu=0, 1, 2,\ldots$ and Fourier transforming
we obtain for Eq (\ref{e8}) \cite{Namiki}   
\begin{equation} \label{e12} \frac{\partial \tilde{q}_i^k(\kappa,s)}{\partial s}=
 -m((\kappa_k)^2+(w_0)^2)\tilde{q}^k_i(\kappa_k,s_k)+\tilde{\eta}_i(\kappa_k,s_k),  
 \end{equation}
 with the following Gaussian constraints (in which we denote 
 the Fourier transforms of $q_i(t,s_k)$ 
and $\eta_i(t,s_k)$ by $\tilde{q}_i(\kappa_k,s_k)$ and 
$\tilde{\eta}_i(\kappa_k,s_k)$ respectively). 
 $$ <\!\tilde{\eta}_i(\kappa_k,s_k)\!>=0, \ \ \ 
<\!\tilde{\eta}_i(\kappa_k,s_k)\tilde{\eta}_j(\grave \kappa_k,\grave s_k)\!>=
 2\delta_{ij}\delta (\kappa_k+\grave \kappa_k)\delta (s_k-\grave s_k) $$
 In the following we write $s_{k-1}$ for $\grave s_k$. 
Solving Eq (\ref{e12}) for $\tilde{q}_i(\kappa_k,s_k)$ one obtains
\cite{Namiki} 
\begin{eqnarray}  && \tilde{q}_i^k(\kappa_k,s_k)=q_0
\exp(-m((\kappa_k)^2+(w_0)^2)s_k)+
\int_0^{s_k}\exp(-m((\kappa_k)^2+ \label{e13} \\ && +(w_0)^2)(s_k- s_{k-1}))
\tilde {\eta}_i(\kappa_k,s_{k-1})ds_{k-1} \nonumber  
\end{eqnarray}
 From the last equation we obtain  the correlation  
$\tilde {P}_{ij_{\tilde
{q}}}(\kappa_k,s_k-s_{k-1})$  
\begin{eqnarray}  
&& \tilde {P}_{ij_{\tilde
{q}}}(\kappa_k,s_k- s_{k-1})=<\!\tilde{q}_i^k(\kappa_k,s_k)\tilde{q}_j^k
(\kappa_k,s_{k-1})\!>= \label{e14} \\ && = \frac{1}
{m((\kappa_k)^2+(w_0)^2)}\exp(-m((\kappa_k)^2+  (w_0)^2)|s_k-s_{k-1}|) 
\nonumber \end{eqnarray} 
Since we want our results to  include a time
dependence we Fourier transform Eq (\ref{e14}) back to obtain 
\begin{eqnarray}  
&& P_{ij_q}(t_k-t_{k-1},s_k-s_{k-1})=\frac{1}{2\pi}\int
d\kappa_k\frac{1}{m((\kappa_k)^2+(w_0)^2)}\exp(i\kappa_k (t_k-t_{k-1})
- \label{e15} \\ && - m((\kappa_k)^2 +(w_0)^2)|s_k-s_{k-1}|) \nonumber \end{eqnarray} 
The former equations (\ref{e12})-(\ref{e15}) are for the subintervals  
$(t_{k-1},t_k)$ and 
$(s_{k-1},s_k)$   so that  for obtaining the corresponding expression for the 
whole intervals $(t_0,t)$ and 
$(s_0,s)$ we use  the following property of correlation functions 
\cite{Klauder} that if $<\!q_i(x_i)q_j(x_j)\!>=<\!q_i(x_i)\!><\!q_j(x_j)\!>$ then 
$$<\!q_1(x_1)q_2(x_2)\ldots q_{2N}(x_{2N}\!>=\prod_{k=0}^{k=N-1}<\!q_{2k+1}(x_{2k+1})
q_{2k+2}(x_{2k+2})\!> $$
Thus, the generalization of Eq (\ref{e15}) to the entire intervals is    
 \begin{eqnarray} && P(q^{(2N-1)},t,s|q^0,t_0,s_0)_{st}=<\!q_0(t_0,s_0)q_1(t_1,s_1)\ldots q_{2N-1}
(t_{2N-1},s_{2N-1})\!>=\nonumber  \\ && = 
(\frac{1}{2\pi})^N\prod_{k=0}^{k=N-1}\int d\kappa_k \frac{1}
{m((\kappa_k)^2+(w_0)^2)}\exp(i\kappa_k (t_{2k+1}- t_{2k})- \label{e16} \\ && - 
m((\kappa_k)^2+(w_0)^2)|s_{2k+1}- s_{2k}|)
 \nonumber \end{eqnarray}
  Eq (\ref{e16})  is  the sought for probability 
  to find at $t$ and $s$ the whole  of the
 ensemble  of $2N$ programmers  having  the same 
  harmonic oscillator code $q^{2N-1}$
 if at the initial  $t_0$ and  $s_0$ they all begin from   
 $q^0$ which is the initial stage at which they begin to write their code. 
 As remarked, the stationary configuration is obtained in  the limit of
 eliminating $s$ so 
equating all its different values  to each other, as required by the SQ
method,  one have  \begin{eqnarray}  
&& P(q^{2N-1},t,s|q^0,t_0,s_0)_{st}=(\frac{1}{2\pi})^N 
\prod_{k=0}^{k=N-1}\int d\kappa_k \frac{e^{i\kappa_k(t_{2k+1}- t_{2k})}}
{m((\kappa_k)^2+(w_0)^2)}
=  \label{e17} \\ && = 
\prod_{k=0}^{k=N-1}\frac{e^{-w_0|t_{2k+1}- t_{2k}|}}{2mw_0} 
 \nonumber \end{eqnarray} 
Note that the elimination of the variable $s$ is obtained by only equating all
its different values to each other without having to take the infinity limit. 
Figure 1 shows the correlation from Eq (\ref{e17}) 
as a function of $t$ for $m=1$ and $w_0=0.4$.  It begins from the  value of 
1.25, 
which corresponds to the assigned values of $m$ and $w_0$, 
then steps  through 
a maximum and vanishes for large $t$.  Figure 2 shows a three-dimensional 
graph of the general 
correlation from Eq (\ref{e16}) as function of  $t$ and $s$ and for the 
same values of $m=1$ and 
$w_0=0.4$ as in Figure 1. Note that for large $s$ the correlation vanishes even at 
those values of $t$ at 
which it attains its maximum in the stationary case of Figure 1.  The values of
$m=1$ and $w_0=0.4$ are typical values used for simulation purposes. \par 
 Note that the stationary state  from Eq (\ref{e17}) 
 have been  obtained  by inserting the harmonic oscillator Langevin 
expression  from Eq (\ref{e12})
into  the action $S$  of  
each  subinterval pair $(t_{k-1},t_k),\   (s_{k-1},s_k), \ \ 1 \le k \le 2N$  
 of each member of the ensemble. 
   This kind of substitution is clearly 
   seen in  Eq (\ref{e10}) which  includes the Langevin relation from
 (\ref{e8}) in each pair of subintervals $(t_{k-1},t_k)$, $(s_{k-1},s_k)$. 
 Note that the substituted expressions of the harmonic oscillator into the
 actions of the subintervals are, of course, not identical since in this case
 the probability to write the same code by all the programmers 
  is trivially unity. 
 As one may realize from Eq (\ref{e16}) the substituted expressions differ by
 $s$ and $t$ and only in the limit that these expressions have the same $s$ and
 $t$ that one finds the same code  shared by all the ensemble members.  
  \par As noted, the ultimate test of any simulation is that it correctly
  describe  the simulated phenomena when the written code is run on the computer
  screen.  Thus,    
   taking into account that  the
classical path integrals discussed here   are formulated 
 in the Euclidean formalism \cite{Roepstorff,Swanson} in which
 the time $t$ is imaginary  we see that the correlation from 
 Eq (\ref{e17}) is almost 
the
same as that of the quantum harmonic oscillator which is \cite{Namiki}
\begin{equation}  \label{e18} \Delta_{quantum}(t_1-t_0,t_3-t_2,\ldots 
  t_{2N-1}-t_{2N-2}) 
  =\prod_{k=0}^{k=N-1}\frac{e^{-iw_0|t_{2k+1}- t_{2k}|}}{4\pi mw_0}
   \end{equation} 
    Thus, as remarked, introducing into the  
    numerical code  of the harmonic oscillator all the detailed description of
    the real one 
      we obtain 
    the correct  simulation as seen from comparing  Eq (\ref{e17}) to 
    (\ref{e18}).    
    Moreover, 
when the remarked 
 substitutions of the harmonic oscillator  relation is  performed in a 
 dense manner over  very 
 short intervals of $t$ and $s$, in which case the substituted expressions 
  are almost
 identical, one may obtain the situation in which all the members of the
 ensemble write exactly the same code   and,  therefore,  
 the correlation  (\ref{e17})    becomes large. In such case $N$ becomes very
 large and may be written  
 as  $N=\frac{t-t_0}{2\delta t}$  where $\delta t$  
  is  the time difference of each of the $2N$ subintervals. Thus,  
  Eq (\ref{e17}) becomes \begin{equation} \label{e19}
  P(q^{2N-1},t,s|q^0,t_0,s_0)_{st}=\frac{e^{-Nw_0\delta t}}{(4\pi mw_0)^N}=(\frac{e^{-w_0\delta t}}{4\pi mw_0})^N
  \end{equation}
  From the last equation  one realizes that  if the
  condition    
  \begin{equation}  \label{e20} e^{-w_0\delta t}=4\pi mw_0, 
\end{equation}   
 is satisfied then the correlation among the ensemble members is maximal because
  each of  them   
 writes  exactly the same code  so   the mentioned probability  
  is unity. 
 Note that in this case not
 only the $s$ intervals tends to zero  but also the $t$'s  as seen
from the former equations. In this case the left hand side of Eq (\ref{e20})
becomes almost  unity  (for not very large  values of $w_0$) 
and thus for having a probability of unity one  have
$w_0=\frac{1}{4\pi}$ (keeping the former value of $m=1$). 
  
\bigskip 
 \pagestyle{myheadings}
\markright{ THE LAMB SHIFT EXAMPLE}
         
\bigskip        \protect
\section{The Lamb shift example}
We see from the former section that substituting the harmonic oscillator
expression into the actions $S$ of the
path integrals \cite{Feynman2,Roepstorff} which are  related to  
the large ensemble of programmers   
 establishes  it  among them in the sense that the
probability to find them writing the same  program  which represents the
harmonic oscillator  is large.  We show this again for the 
 example of  a two-state electron which emits a photon and then reabsorbs it 
 where the total
energy is not conserved. We assume, as for the harmonic oscillator example, 
 that there are a large number of different
numerical codes  of this process which  reflect the large number of different 
ways  which 
lead to the same  simulation. Thus, as for the harmonic oscillator case,  we
introduce  a large ensemble of 
programmers   all of them want to
simulate this electron-photon interaction upon their computer screens. 
As in the
former section the differences among the written programs  may
be related to the different values of the extra variable $s$.  And,  as for the
harmonic oscillator case, the equilibrium state   
is obtained when all the different values of $s$ are equated to each other and
taken to infinity. 
\par 
We subdivide  the intervals $(s_0,s)$ and $(t_0,t)$, during which this process occurs,
 into a large number of 
subintervals $(s_0,s_1)$, $(s_1,s_2)$, \ldots $(s_{N-1},s_N)$ and $(t_0,t_1)$, 
$(t_1,t_2)$, \ldots, $(t_{N-1},t_N)$  and formulate the appropriate
expression for the described electron-photon interaction over the representative 
subintervals  
$(t_{k-1},t_k)$ and $(s_{k-1},s_k)$. We  calculate the probability to
find the ensemble of programmers writing the same code which truly   
   simulates  
the  remarked electron-photon
interaction. We find it better to discuss  now this probability  
 using  the Fokker-Plank equation 
\cite{Namiki,Risken}. 
 That is, we begin from  \cite{Namiki,Risken}
 \begin{equation} \frac{\partial P(q^k,t_k,s_k|q^{k-1},t_{k-1},s_{k-1})}{\partial s}=
F(q^k)P(q^k,t_k,s_k|q^{k-1},t_{k-1},s_{k-1}),    \label{e21}  \end{equation} 
where  $P(q^k,t_k,s_k|q^{k-1},t_{k-1},s_{k-1})$ is given by Eq (\ref{e10}) 
and denotes  the probability to find the relevant ensemble of programmers  
 at $t_k$ and $s_k$ having the code  $q^k$ if at the former $t_{k-1}$ and   
 $s_{k-1}$   they have  the code 
$q^{k-1}$. Note that, as in the harmonic oscillator example,  the word 
code does not necessarily means
the complete program of this process, that is,   even a very small part
of it is called code.  
The operator $F(q^k)$ is  \cite{Namiki} \begin{equation} \label{e22} 
F(q^k)= \frac{1}{2\alpha}H(q^k,\pi^k), \end{equation} where $H$ and 
$\pi^k$ are the ``stochastic'' Hamiltonian and momentum respectively and
$\alpha$ is as discussed after  Eq (\ref{e3}).  The momentum operator $\pi^k$ is defined as in
quantum mechanics \cite{Namiki} $\pi^k=-2\alpha \frac{\partial }{\partial q^k}$, 
and  its commutation with the   operator $q^i$  satisfy 
\cite{Namiki} $[\pi^k,q^i]= 2\alpha \delta_{ki}$, where all one have to do in
order to obtain the quantum regime is to set  \cite{Namiki}
$\alpha=\frac{i\hbar}{2}$. From the former  relations one may
develop,  as
has been done in \cite{Namiki}, an operator formalism similar to that of quantum
mechanics, especially, the corresponding ``Schroedinger'', ``Heisenberg'' and
``interaction'' pictures.  \par 
Using the former discussion  we find the conditional probability
 to find at $s$ and $t$ the ensemble having the code  $q^N$ if at the initial 
 $s_0$
 and $t_0$ they have the code  $q^0$. That is, one can write  
 this probability in the
 ``interaction'' picture for the 
 intervals $(t,t_0)$, and  $(s,s_0)$   \cite{Namiki,Haken} 
 \begin{eqnarray} && P^I(q^N,t_N,s_N|q^0,t_0,s_0)=P^I(q^0,t_0,s_0)+
 \label{e23} \\ && + \int
 F^I(q^N)P^I(q^{N-1},t_{N-1},s_{N-1}|q^0,t_0,s_0)dq^{N-1}, \nonumber 
  \end{eqnarray} 
 where $P^I(q^0,t_0,s_0)$ is the probability to find the system at the initial 
 $t_0$ and $s_0$ with  the code $q^0$  which probably contains at the initial
 stage only a few bits  and the superscript $I$
 denotes that we consider the "interaction" picture.  Note that 
  $q$ depends upon $s$ and $t$ so  the integral with respect to
 $q$ is, actually, a double one  over $s$ and $t$. Substituting,  in a
 perturbative manner \cite{Feynman2}   for 
 $P^I(q^{N-1},t_{N-1},s_{N-1}|q^0,t_0,s_0)$   on the right hand side of Eq
 (\ref{e23})  one
 obtains \begin{eqnarray} &&
 P^I(q^N,t_N,s_N|q^0,t_0,s_0)= 
 \sum_{n=0}^{n=\infty}\frac{1}{n!}\int_{q^0}^{q^N}dq^1\int_{q^0}^{q^N}dq^2
 \ldots \int_{q^0}^{q^N}dq^N T (F^I(q^1)F^I(q^2)\ldots  \nonumber \\ && \ldots 
 F^I(q^N)P^I(q^0,t_0,s_0)
 =P^I(q^0,t_0,s_0)+\int_{q^0}^{q^N}dq^1F^I(q^1)P^I(q^0,t_0,s_0)+ 
 \label{e24} \\ && +
 \int_{q^0}^{q^N}dq^2 \int_{q^0}^{q^2}dq^1F^I(q^1)F^I(q^2) P^I(q^0,t_0,s_0)+ 
 \ldots 
  \int_{q^0}^{q^N}dq^{N-1} \int_{q^0}^{q^{N-1}}dq^{N-2} 
 \ldots \nonumber \\ && \ldots \int_{q^0}^{q^1}F^I(q^1)F^I(q^2) \ldots F^I(q^N)P^I(q^0,t_0,s_0) \nonumber 
 \end{eqnarray}
 We, now, follow the same rules in \cite{Haken}, except for the introduction
 of the variable $s$, for representing the electron and photon before and after
 their interaction. The extra variable $s$ is introduced into the relevant
 quantities so that in the limit of $s \to \infty$, as required in the SQ
 method \cite{Parisi,Namiki},   the known expressions \cite{Haken}
 which   represent the electron and photon are obtained.    
 We note first that  the probability $P^I$    
  is no other than   the {\it state} of the system \cite{Namiki} 
  (as in quantum mechanics the system states of the SQ theory have 
   also a probabilistic 
  character). Thus,
 in the former equations we may assign to the initial $s_0$ and $t_0$ 
 the value of zero
 and refer to $P^I(q^0,t_0=0,s_0=0)$ as the initial state of the  
 ensemble system.  This initial common state denotes, as for the harmonic
 oscillator case discussed in the former section, the common starting point of
 all the ensemble programmers  which  simulate 
  the electron-photon interaction.  \par 
 As remarked, the electron  is assumed to have 
 two different states so  that at $t_1$ and $s_1$ it was at the higher state 2
 from which  it descends to the lower one 1 through emitting a photon. Then at
 $t_2$ and $s_2$ it reabsorbs the photon and returns to state 2 as schematically
 shown at the left hand side of Figure 3. The incoming electron and the emitted 
 photon at $t_1$ and $s_1$ may be represented by 
$e^{-i\epsilon_2t_1}+e^{-i\epsilon_2s_1(1-i\delta)}$  
   and    $e^{-iw_{\lambda}t_1}+e^{-iw_{\lambda}s_1(1-i\delta)}$ respectively,      
   where $\delta$ is an 
infinitesimal satisfying $\delta \cdot \infty=\infty$, and $\delta \cdot c=0$, 
 ($c$ is a constant) \cite{Mattuck}. This is done   
  so that in the equilibrium configuration, 
  which is obtained in  the SQ theory when $s \to \infty$, 
   the terms in $s$
 vanish  as required 
 \cite{Parisi,Namiki}  and one remains only with those in $t$ as in 
 \cite{Haken}. 
The outgoing electron after emission at $t_1$ and $s_1$
 may be  represented by the plane wave $e^{i\epsilon_1t_1}+
 e^{i\epsilon_1s_1(1+i\delta)}$ where the $\delta$ has the same meaning as 
 before.  At the reabsorption stage at $t_2$
  and $s_2$ the  electron is represented, before absorbing the photon, 
    by $e^{-i\epsilon_1t_2}+e^{-i\epsilon_1s_2(1-i\delta)}$ and after the
    absorption by 
    $e^{i\epsilon_2t_2}+e^{i\epsilon_2s_2(1+i\delta)}$.  The photon is
    represented at the reabsorption stage by  
    $e^{iw_{\lambda}t_2}+e^{iw_{\lambda}s_2(1+i\delta)}$.  Also, the emission 
    itself, 
    denoted by the vertex in
  Figure 3, may be represented, as in the quantum analog \cite{Haken}, by
  $g_{\lambda_s}$ and the reabsorption by  $g^+_{\lambda_s}$, where an explicit
  expressions for   $g_{\lambda_s}$ and $g^+_{\lambda_s}$ may be obtained in an
  equivalent manner to their quantum analogs (see \cite{Haken}), but these
  expression are not required for the discussion here. Thus, since the final
  state at $t$ and $s$ after the reabsorption of the photon is the same as the 
  initial one before its emission we may  write for the relevant $P^I$ at
  the end of the whole  process of emission and reabsorption \cite{Haken} \begin{equation}
  \label{e25}  P^I(q^N,t_N,s_N|q^0,t_0,s_0)=P^I(q^0,t_0,s_0)+C(t,s)P^I(q^0,t_0,s_0) 
  \end{equation}  The coefficient $C(t,s)$ denotes the mentioned evolution from 
   the initial state $P^I(q^0,t_0,s_0)$ 
   to the final one $P^I(q^N,t_N,s_N|q^0,t_0,s_0)$  
  and   is found as in  
   \cite{Haken} which  discusses   the same process  in quantum terms  
   (without using the variable
  $s$). We first note    that the entire  interaction  of
  (emission$+$reabsorption) 
  in the variables $t$ and $s$,  which is described  after Eq
  (\ref{e24}), may be written as a sum of two separate terms 
  $P(t)$ and $P(s)$   each of them involves 
   only one variable. These two terms are      
   \begin{eqnarray} &&  P(t)=
 g_{\lambda_s}g^+_{\lambda_s}
\int_0^{t_2}\exp(i(\epsilon_1-w_{\lambda}-\epsilon_2)t_1)dt_1
\int_0^t\exp(i(\epsilon_2+w_{\lambda}-\epsilon_1)t_2)dt_2 \nonumber \\
  && P(s)=
 g_{\lambda_s}g^+_{\lambda_s}
\int_0^{s_2}\exp(i(\epsilon_1+i\delta(\epsilon_2+\epsilon_1+
w_{\lambda})-w_{\lambda}-\epsilon_2)s_1)ds_1 \cdot \label{e26} \\
&& \cdot \int_0^s\exp(i(\epsilon_2+i\delta(\epsilon_2+\epsilon_1+w_{\lambda})+
w_{\lambda}-\epsilon_1)s_2)ds_2,   \nonumber 
 \end{eqnarray} 
 where we have set, as remarked,  $s_0=t_0=0$.  Each of the 
 two expressions $P(t)$ and
 $P(s)$   is, actually,     an account of the whole process of emision and
 reabsorption, as discussed after Eq (\ref{e24}), in the respective 
 variables $t$
 and $s$.   
    $C(t,s)$ from Eq (\ref{e25}) is given by  the sum $P(t)+P(s)$ so that 
    in the equilibrium
    state obtained in the limit in which all the values of $s$ are
    equated to each other the term $P(s)$ vanishes and remains only the term
    $P(t)$ as should be \cite{Haken}. This is because  we have already equated
    the initial $s_0$ to zero so for equating all the other $s$'s  to 
    each other
    one have to set also the other values of $s$ equal to zero which causes
    $P(s)$ to vanish (see the second of Eqs (\ref{e26})). Thus  
 \begin{eqnarray} && C(t,s)=P(t)+P(s)=
\sum_{\lambda_s}g_{\lambda_s}g^+_{\lambda_s}\int_0^tdt_2\frac{(\exp(i(\epsilon_1-
\epsilon_2-w_{\lambda})t_2)-1)}{i(\epsilon_1-\epsilon_2-w_{\lambda})} \cdot
\nonumber \\ && \cdot 
\exp(i(\epsilon_2+w_{\lambda}-\epsilon_1)t_2)+
 \sum_{\lambda_s}g_{\lambda_s}g^+_{\lambda_s} \cdot \label{e27}  \\ && \cdot 
\int_0^sds_2\frac{(\exp(i(\epsilon_1-
\epsilon_2-w_{\lambda}+i\delta(\epsilon_2+\epsilon_1+w_{\lambda}))s_2)-1)}
{i(\epsilon_1-\epsilon_2-w_{\lambda}+
i\delta(\epsilon_2+\epsilon_1+w_{\lambda}))}
\exp(i(\epsilon_2-\epsilon_1+w_{\lambda}+ \nonumber \\ 
&& +i\delta(\epsilon_2+\epsilon_1+w_{\lambda}))s_2)
= 
\sum_{\lambda_s}
\frac{g_{\lambda_s}g^+_{\lambda_s}}{i(\epsilon_1-\epsilon_2-w_{\lambda})}[t-
(\frac{(\exp(i(\epsilon_2-
\epsilon_1+w_{\lambda})t)-1)}{i(\epsilon_2-\epsilon_1+w_{\lambda})})]
+ \nonumber \\ && +
\sum_{\lambda_s}
\frac{g_{\lambda_s}g^+_{\lambda_s}}
{i(\epsilon_1-\epsilon_2-w_{\lambda_s}+
i\delta(\epsilon_2+\epsilon_1+w_{\lambda}))}
[\frac{(e^{i2\delta (\epsilon_2+\epsilon_1+w_{\lambda})s}-1)}
{i2\delta(\epsilon_2+\epsilon_1+w_{\lambda})}- \nonumber \\ && 
-(\frac{(\exp(i(\epsilon_2-
\epsilon_1+w_{\lambda}+i\delta(\epsilon_2+\epsilon_1+w_{\lambda}))s)-1)}
{i(\epsilon_2-\epsilon_1+w_{\lambda_s}+
i\delta(\epsilon_2+\epsilon_1+w_{\lambda}))})] \nonumber \end{eqnarray}
 The first quotient  in the square parentheses of the second sum, which is of the
 kind $\frac{0}{0}$, may be evaluated,   using  L'hospital theorem
 \cite{Pipes},  to obtain for it the result of $s$ so that Eq (\ref{e27}) becomes
  \begin{eqnarray} && C(t,s)=P(t)+P(s)=
  \sum_{\lambda_s}
\frac{g_{\lambda_s}g^+_{\lambda_s}}{i(\epsilon_1-\epsilon_2-w_{\lambda})}[t-
(\frac{(\exp(i(\epsilon_2-
\epsilon_1+w_{\lambda})t)-1)}{i(\epsilon_2-\epsilon_1+w_{\lambda})})]
+  \nonumber \\ && + \sum_{\lambda_s}
\frac{g_{\lambda_s}g^+_{\lambda_s}}
{i(\epsilon_1-\epsilon_2-w_{\lambda_s}+i\delta(\epsilon_2+\epsilon_1+w_{\lambda}))}
[s-  \label{e28} \\ && - (\frac{(\exp(i(\epsilon_2-
\epsilon_1+w_{\lambda}+i\delta(\epsilon_2+\epsilon_1+w_{\lambda}))s)-1)}
{i(\epsilon_2-\epsilon_1+w_{\lambda_s}+i\delta(\epsilon_2+\epsilon_1+w_{\lambda}))})] \nonumber 
 \end{eqnarray} 
The last expression for $C(t,s)$ yields terms of several kinds among them those
which  are proportional to $t$ and
$s$,   others which  are oscillatory in these variables,  and also 
constant terms. Thus,  
 for large $t$ and $s$ the oscillatory as well as the constant terms may
be neglected compared to $t$ and $s$ as in the analogous quantum discussion  of the
same process \cite{Haken}. Substituting the resulting  expression in Eq
(\ref{e25}) one obtains \begin{equation} \label{e29}
P^I(q^N,t_N,s_N|q^0,0,0)=P^I(q^0,0,0)(1+C(t,s))=P^I(q^0,0,0)(1+it\Delta
\epsilon_{\lambda}+is\Delta \epsilon_{\lambda_s}), \end{equation} 
where, 
\begin{equation} \label{e30} \Delta \epsilon_{\lambda}=\sum_{\lambda_s}\frac
{g_{\lambda_s}g^+_{\lambda_s}}{\epsilon_2-\epsilon_1+w_{\lambda}}, \ \ \ 
\Delta \epsilon_{\lambda_s}=\sum_{\lambda_s}\frac
{g_{\lambda_s}g^+_{\lambda_s}}{\epsilon_2-\epsilon_1+w_{\lambda_s}-
i\delta(\epsilon_2+\epsilon_1+w_{\lambda})}
\end{equation} 
The result in Eq (\ref{e29}) is only for the first-order term in Eq (\ref{e24}).
 If all the higher order terms of this process are taken
into account  one obtains, analogously to the quantum analog (in which the
variable $s$ is absent), the result \begin{eqnarray} && P^I(q^N,t_N,s_N|q^0,0,0)=
P^I(q^0,0,0)(1+C(t,s))=P^I(q^0,0,0)(1+(it\Delta \epsilon_{\lambda}+ \nonumber \\
&&+ \frac{1}{2!}
(it\Delta \epsilon_{\lambda})^2+ \ldots +  \frac{1}{n!}
(it\Delta \epsilon_{\lambda})^n+\ldots)+ (is\Delta \epsilon_{\lambda_s}+
\frac{1}{2!}
(it\Delta \epsilon_{\lambda_s})^2+ \ldots  \label{e31} \\ && \ldots + \frac{1}{n!}
(it\Delta \epsilon_{\lambda_s})^n+\ldots) 
 =P^I(q^0,0)(e^{it\Delta \epsilon_{\lambda}}+e^{is\Delta \epsilon_{\lambda_s}} -1)
\nonumber \end{eqnarray} 
The right hand side of Figure 3 shows the diagram of the fourth order term 
of this process. 
Now, as required by the SQ theory, the  stationary situations are obtained in the limit
of eliminating the extra variable $s$ which is done by equating all the $s$
values to each other and taking to infinity.  Thus,  since, as remarked,  we have  
 equated the initial $s_0$ to zero we must  equate all the
other $s$ values to zero. That is, the stationary state is \begin{equation}
\label{e32} \lim_{s\to 0}P^I(q^N,t_N,s_N|q^0,0,0)= \lim_{s\to 0}P^I(q^0,0,0)
(e^{it\Delta \epsilon_{\lambda}}+e^{is\Delta \epsilon_{\lambda}}-1)=
P^I(q^0,0)e^{it\Delta \epsilon_{\lambda}} \end{equation}
The last result is the one obtained in quantum field theory \cite{Haken} 
for  the
same interaction of (emission$+$reabsorption). The
quantity $\Delta \epsilon_{\lambda}$,  given by the first of Eqs (\ref{e30}), 
has the same form also in the quantum version \cite{Haken,Mahan},   
 where it is 
termed the energy
shift.  This   shift have  
been experimentally demonstrated in the quantum
field theory for the  case of a real
many-state  particle in the famous Lamb shift of the Hydrogen atom  
\cite{Haken,Mahan,Lamb}. \par 
 As remarked,  this  electron-photon interaction is discussed in the literature 
\cite{Haken,Mahan,Lamb}  without using
any extra variable and the  result at  the right hand side of Eq 
(\ref{e32}) is obtained.  This    result  have been  obtained here  by
discussing the simulation process of the electron-photon interaction, that is, 
by  using  
 the
extra variable $s$ in the limit of equating all its values.  In other words, as
for the harmonic oscillator case, introducing the expression of the detailed
electron-photon interaction into all the relevant codes (into 
 the subintervals of $s$ and $t$ of all the programmers)
yields a correlation among them which truly represents the corelation of the
real interaction.  
Thus, as for the harmonic oscillator example in
which the use of the SQ method and the extra variable $s$ lead to the real
known quantum correlation (compare the two Equations (\ref{e17}), (\ref{e18})), 
so also here we obtain, using the same method and extra variable, the known
expression for the probability $P^I(q^N,t|q^0,t_0)$. 
 Also,  the writing stage of
the relevant programs, which is characterized by the different versions of 
the code, is represented  
by the last result of Eq 
 (\ref{e31})  (the analogous writing stage of the harmonic oscillator example
 is given by Eq (\ref{e16})). 
 The equilibrium stage  corresponds to the case where all the values of $s$
 are equated to each other as seen from  Eq (\ref{e32}) in which
 all the  $s$'s, including the initial $s_0$ (see the discussion after
 Eqs (\ref{e24}) and (\ref{e31})), are assigned the value of zero. The 
 analogous expression for the harmonic oscillator example is given by Eq
 (\ref{e17}). The corresponding similarities between the correlations for the
 two cases are, as remarked, the ultimate tests which show that the
 relevant simulations truly represent the simulated phenomena.   \par  

   \protect \section*{\bf Concluding Remarks \label{sec4}} \noindent
We have discussed  the code-writing stage of  the numerical  simulations  of 
real phenomena using 
physical terms and terminology.  The method applied  
 for discussing this code-writing stage  
 is 
 the stochatic quantization  method  of Parisi-Wu-Namiki \cite{Parisi,Namiki}
 where  an extra variable is
introduced  that takes account of an assumed stochastic process (in
this variable) which allows  a large number of 
possible different
 behaviours of the  system. The equilibrium configuration is
obtained \cite{Parisi,Namiki}  when this variable is eliminated through 
equating all its different
values to each other and taking to infinity.    
This equating of all the possible $s$ values
to each other introduces an element of repetitions  
of the same process through     
 which the
system is stabilized and brought to its  equilibrium configuration. \par
We discuss the system  of a large ensemble of programmers which all try to 
 simulate the same phenomena  by writing the appropriate code 
 which naturally  will 
 not be the same for all of them.  
 The mentioned equating of all 
the $s$ values  means that  all of them write  the same code 
  and this constitutes   
a repetition of  of writing  the same program so that  the probability 
to 
see the same numerical simulation  is obviously large.   \par 
We note that obtaining numerical equilibrium configuration  through running 
a large 
number of 
times  the related  code upon the computer screen
 is the main characterstic of many  
 simulations processes especially those concerned with finding numerical
solutions of  physical situations. For example, any one who try  to numerically
solve any differential equation which  governs the evolution of some physical
system knows, as shown in the following, that the solution suggested by the
computer is obtained only after repeatedly updating the given differential
equation.   Better statistics is obtained when the
number of iterations grows since this increases also the number of samples. 
The advantage 
of these repetitions is clearly seen for the case of simulating the long 
range correlation functions \cite{Parisi2,Namiki2} 
for which 
 the conventional Monte Carlo simulation methods  
  to numerically simulate  them, using path 
integrals,   fails \cite{Namiki2}.  
It has been shown  explicitly by Parisi  \cite{Parisi2},   and Namiki 
{\it et al}  \cite{Namiki2},  
using SQ methods,   that the following two point connected correlation function 
\cite{Roepstorff} 
 \begin{equation} \label{e47} <\!q_1q_l\!>=\frac{d}{dh}(\frac{\int
 d[q]q_le^{-S_h}}{\int d[q]e^{-S_h}})|_{h=0}=
 \frac{1}{h}(<\!q_l\!>_h-<\!q_l\!>)
+o(h),  \end{equation} where the action $S_h=S-hq_1$ involves a small external
 additional term $-hq_1$, may be solved by replacing the ensemble averages with
 and without the small external source $<\!q_l\!>_h$ and  $<\!q_l\!>$
 respectively by the time averages calculated from the following Langevin
 equations
 \begin{equation} \label{e48} \dot {q_l}=-\frac{\partial S}{\partial q_l}+\eta_l
\end{equation} 
 \begin{equation} \label{e49} \dot {q_l}=-\frac{\partial S_h}{\partial q_l}+
 \tilde {\eta_l}
 \end{equation} 
The $\eta$ and  $\tilde {\eta_l}$ are independent and assumed to satisfy 
$$<\!\eta \!>=<\!\tilde {\eta_l}\!>=0, \ \ \ \ <\!\eta_l(t)\eta_m(s)\!>=
<\!\tilde \eta_l(t)\tilde \eta_m(s)\!>=2\delta_{lm}\delta(t-s)$$
Solving the right hand side of Eq (\ref{e47})
generally results in a large statistical error  
\cite{Parisi2,Namiki2},    so Parisi  \cite{Parisi2}
uses the same random
forces in Eqs (\ref{e48}), (\ref{e49}), that is, $\eta=\tilde {\eta}$ which
reduces considerably the statistical error as shown in \cite{Namiki2} 
(see the 
Appendix there). Thus,  in order to be able
to simulate and obtain the long range correlation functions one must  equate
to each other not only all the different values  of $s$ which are related to 
 $\eta$ and  $\tilde {\eta}$ 
but also to equate   $\eta$ to  
$\tilde {\eta}$ so that in the stationary limit 
 one obtains, as remarked, the sought-for simulations.   But as noted by 
Namiki {\it et al} \cite{Namiki2} 
the last method, 
 although works well for the fixed potential and the $o(4)$ model
 \cite{Roepstorff,Swanson},   
 breaks down when one uses it to obtain the long range correlation function for the
 $o(3)$ model possibly due to its large degree of nonlinearity. This situation
is avoided in \cite{Namiki2} by initiating a new round of repetitions where
each one begins from the final configuration of the former. That is, in order to
improve the statistical results one have first, as remarked,  to increase the number of
samples which is obtained by  parallel updating of Eqs (\ref{e48}),
(\ref{e49})  without and with the external source 
respectively using the same random forces for $\eta$ and $\tilde \eta$.  
These steps which are sufficient, as remarked,  for the fixed
potential and the $o(4)$ models end in a breakdown of the simulation  for the 
$o(3)$ model when the updating  process continues.   Thus, one must 
\cite{Namiki2}  stop this 
updating  
before break-down occurs and restart the whole procedure from switching again
the external source and updating Eqs  (\ref{e48}), 
 (\ref{e49}) starting from the last stopped configuration as the  initial
 one of the new round of updating. In other words, by only repeating the switching and 
the updating  process one obtains, numerically, the sought-for long range correlation 
functions  for the $o(3)$ model. \par
Moreover, it has been  shown
\cite{Misra,Itano,Aharonov}, 
that these repetitions 
not only lead to numerical stabilization but when they are really performed  (not just  
through clicking upon the computer keyboard) lead to a real 
stabilization. This 
phenomenon, termed the  Zeno effect, have been validated both 
theoretically \cite{Misra,Aharonov},  
  and experimentally 
\cite{Itano}.   The  main
characteristic of this effect is the preservation, through a large number of 
repetitions of the same 
measurement, of an initial state of the sytem  
\cite{Misra,Itano}, 
or {\it guiding} 
it through a prescribed path of evolution, 
from a large number of possible paths  \cite{Aharonov}. \par 
The principle of repetition have  been shown in Section 3 for the Harmonic oscillator example 
(see Eqs (\ref{e16})-(\ref{e17}) and the discussion there) where we see that when 
 the same  version of program is shared among the ensemble members  then the probability to find 
the same simulation of the harmonic oscillator  in all the screens is large.  
 Moreover, 
when the number of times of performing this simulation,   which 
is related to the number of subintervals (see the discussion after Eq 
(\ref{e16})) 
of  the finite total paths in $s$ and $t$  of each member of the ensemble,   
 becomes large so that the duration 
of each is small then the remarked probability is unity 
(see Eqs (\ref{e19})-(\ref{e20})).  This is so since  each member of the 
ensemble 
have  in this case exactly the same code related to the 
Harmonic oscillator  and the correlation among them is, therefore, maximal. 
The same state of affairs have been found also for the energy shift example
discussed in Section 4. In this case the required correlation is obtained
through summing the relevant Feynman diagram to all orders.  
This influence  of repeating 
 the same experiment a large number of times  
  have been shown to be effective also for classical systems  
  \cite{Bar01}.

\newpage
\bigskip 
\bibliographystyle{}

\begin{thebibliography}{}
\bigskip \parindent 1 in 

\bibitem{Feynman1}  
          R. P.  Feynman, 
			  {\rm ``Simulating Physics with computers''}
			  Int. Jour. Theor. Phys, {\bf 21}, Nos.
                          $\frac{6}{7}$ (1982); 
          R. P.   Feynman,  {\it Found. Phys}, {\bf 16}, 507-531 (1986);  
  C. H.   Bennett, Int. J. Theor. Phys,
				{\bf 21}, 905-940 (1982). 
\bibitem{Naylor}    T. H.   Naylor,  and J. M. Finger,  
                   {\it Management Science B}, 
                        {\bf 14}, N0. 2, 92-101 (1967). 
			
 \bibitem{Kleindorfer}    G. H.   Kleindorfer  and R. Ganeshan, in {\rm ``1993 
                               winter simulation conference proceedings''}, pp. 50-57 ,
                               IEEE, New york  (1993). 
\bibitem{Haken1}   H. Haken, {\it Rev. Mod. Phys}, {\bf 47}, 67 (1975); H.
Haken, Ed,  {\it 
Cooperative
Effects: Progress in Synergetics},  Amsterdam, North-Holland 
(1974);  H. Haken,  {\it Synergetics: Nonequilibrium Phase Transitions and Self 
Organization in
Physics, Chemistry and Biology''},  Springer,  Berlin (1978).		  		       			  
			       
 \bibitem{Lax}   P. D.   Lax and R. S.
                       Phillips, 
		        {\it Scattering Theory}, 
		       Academic, New York (1967).
  \bibitem{Horwitz}    L. P.    Horwitz and  C. Piron, {\it Helv. Phys. Acta}, {\bf 66}, 
                       694 (1993).
 \bibitem{Parisi}     G.   Parisi  and Y. Wu, {\it Sci. Sin}, {\bf 24}, 483
			  (1981); 
                       G.  Parisi, {\it Nuc. Phys}, {\bf B180}, [FS2], 
		       378-384 (1981);   E.    Nelson, 
			       {\it Quantum Fluctuation}, 
			       Princeton
                           University, New Jersey (1985);  
			     E.  Nelson,
			   {\it Phys. Rev A}, {\bf 150}, 1079-1085 (1966). 
   
  \bibitem{Namiki}     M.   Namiki, 
		 {\it Stochastic Quantization}, 
		 Springer, Berlin (1992). 

\bibitem{Coffey}    W.  Coffey, 
			  {\it The Langevin Equation}, 
			 Singapore: 
                               World Scientific (1996). 			   
\bibitem{Risken}    H.    Risken, 
			    {\it The Fokker-Plank Equation}, 
			   Springer (1984).
 \bibitem{Kannan}	  D.   Kannan, 
		   {\it An Introduction to Stochastic Processes}, 
		   Elsevier, 
                       North-Holland (1979); 
                   L. C.    Rogers and D. Williams,
		       {\it Diffusions, Markov
		       Processes and Martingales}, 
		       $2^{nd}$ edition,  Wiley   (1987); 
                  J. L.     Doob, 
		       {\it Stochastic Processes}, 
		       Wiley,
                    New York (1953). 
 \bibitem{Feynman2}	 R. P.   Feynman, {\it Rev. Mod. Phys},{\bf 20}, 2, 367 (1948);  
   R. P.  Feynman and A. R. 
                                Hibbs,  
				 {\it Quantum Mechanics and Path Integrals},   
				McGraw-Hill (1965).
 \bibitem{Haken} 	 H.  Haken,  
			{\it Light},  Vol 1,
			North-Holland 
                                  (1981). 

\bibitem{Mahan}     G.     Mahan, 
			{\it Many Particle Physics},  
			      $2^{nd}$ edition,  
			Plenum, New York (1993); C.   Enz, 
			 {\it   A Course
			on Many Body Theory Applied to Solid State Physics}, 
			 World Scientific (1992).
 \bibitem{Lamb}   W. E.   Lamb, Jr.  and M. Sargent,
		       {\it Laser Physics},		  
		  Addison-Wesley,  Advanced Book Program (1974);  
		       W. E.   Lamb, 
			{\it The Interpretation of Quantum Mechanics},
		       Jr., Rinton Press (2001);  				 				 		      
                       T. W.      Hansch, I. S. Shahin and A. L. Schawlow, 
		       {\it Nature}, {\bf
                      235}, 63 (1972); 
                      T. W.    Hansch, A. L. Schawlow and P.
		      Toschek, {\it IEEE J. Quant. Electr}. QE-8, 802 (1977). 
  

 \bibitem{Roepstorff}  G.   Roepstorff, 
			   {\it Path Integral Approach to Quantum Physics},
			  Springer-Verlag (1994); 
\bibitem{Masao}                D.    Masao, {\it J. Physics A: Math. Gen}, {\bf 9}, 1465-1477 
                        (1976); 		
			D.  Masso, {\it J. Physics A: Math. Gen}, {\bf 9}, 
                       1479-1495 (1976). 
\bibitem{Mattuck}   R. D.   Mattuck, 
			{\it A Guide to Feynman Diagrams in the Many Body 
                       Problem}, $2^{nd}$ edition, 
			McGraw-Hill
		        (1967). 
		       
\bibitem{Klauder}  J. R.    Klauder  and E. C.
                      G. Sudarshan, 
		     {\it  Fundamentals of Quantum Optics}
		      W. A. Benjamin (1968). 
\bibitem{Swanson}   M.    Swanson, 
			{\it  Path Integrals and Quantum Processes}, 
			Academic   (1992). 
 \bibitem{Pipes}   A. L.   Pipes, 
			 {\it Applied Mathematics for Engineers and
                          Physicists},
			 $2^{nd}$  edition, McGraw-Hill  (1958). 
 \bibitem{Parisi2}  G.    Parisi, {\it Nuc. Phys}, {\bf B205}, [FS5], 337-344, (1982). 
  \bibitem{Namiki2}  M.   Namiki  et al, {\it Prog. Phys}, {\bf 76}, 501-511, (1986); 
           M.   Namiki  et al, {\it Prog. Phys}, {\bf 73}, 186-196, (1985). 
    
 \bibitem{Misra}      B.   Misra and E. C. Sudarshan, J. Math. Phys,{\bf 18}, 756 
                (1977); D.    Giulini, E. Joos,  C. Kiefer, J. Kusch,
            I. O. Stamatescu and H. D. Zeh, 
	    {\it Decoherence  and the Appearance of a Classical World in 
            Quantum Theory}, 
	    Springer-Verlag (1996);  S.    Pascazio and Mikio Namiki, 
	    {\it Phys. Rev A }, 
            {\bf 50},  6, 4582  (1994);  A.    Peres, {\it Phys. Rev D }, {\bf 39},  10 
	    2943  
            (1989); 
                A.    Peres and A.  Ron, {\it Phys. Rev A},   {\bf 42},  9,  
		5720  
	                (1990); R. A.  Harris  and 
	              L. Stodolsky, {\it J. Chem. Phys}, 
                   {\bf 74}, 4, 2145  (1981);  M.     Bixon, {\it Chem. Phys}, 
                       {\bf 70}, 199-206 (1982);  M.   Simonius,  
            {\it Phys. Rev. Lett}, {\bf 40},  15, 980-983  (1978). 

 \bibitem{Itano}  W. M.   Itano, D. J. Heinzen, J. J. Bollinger, 
	    and 
            D. J. Wineland, {\it Phys. Rev A}, {\bf 41}, 2295-2300  (1990);  
    	  A. G.    Kofman and G. Kurizki, {\it Phys. Rev A}, {\bf 54}, 
	  3750-3753
	    (1996);  G.     Kurizki, A. G. Kofman and V. Yudson, {\it Phys. Rev A}, {\bf
	    53}, R35 (1995);  S. R.   Wilkinson, C. F. Bharucha, M. C. Madison, P.
	        R. Morrow, Q. Niu,  B. Sundaram and M. G. Raisen, {\it Nature}, {\bf 387},
	    575-577   (1997);   R. J.    Cook, 
	                 {\it Physica 
                         Scripta T}, {\bf 21}, 49-51 (1988). 
   
\bibitem{Aharonov}	     Y.    Aharonov and M. Vardi, {\it Phys. Rev D},  
                    {\bf 21}, 2235  
                    (1980); P. Facchi, A. G. Klein, S. Pascazio and L. Schulman,
		    {\it Phys. Lett A}, {\bf 257}, 232-240 (1999).  			     

\bibitem{Bar01}   D. Bar,   Phys. Rev.  E, {\bf 64}, No: 2, 026108/1-10 
                   (2001);   
              D.    Bar, Physica A, {\bf 292}, 494-508 (2001).	


\end{thebibliography}

\newpage

\begin{figure}[hb]
\centerline{
\epsfxsize=3in
\epsffile{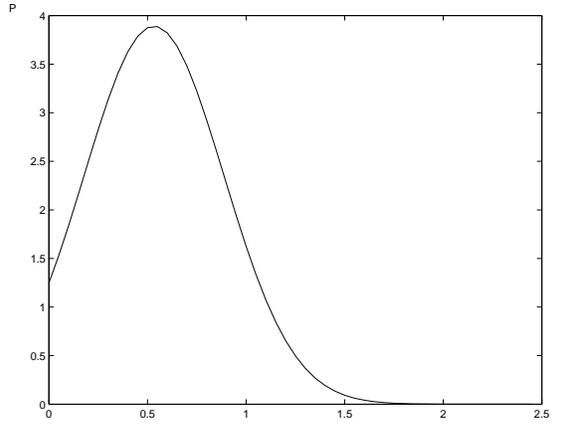}}
\caption[fegf1]{The harmonic oscillator correlation function from Eq (\ref{e17})
as a function of the time $t$ for the values of $m=1$ and initial eigenvalue of
$ w_0=0.4$.  It begins from an initial value of 1.25 (which corresponds 
to the remarked values of $m$ and $w_0$),   proceeds
to a maximum value from which it descends to zero for large $t$. } 
\end{figure}

\begin{figure}[hb]
\centerline{
\epsfxsize=3in
\epsffile{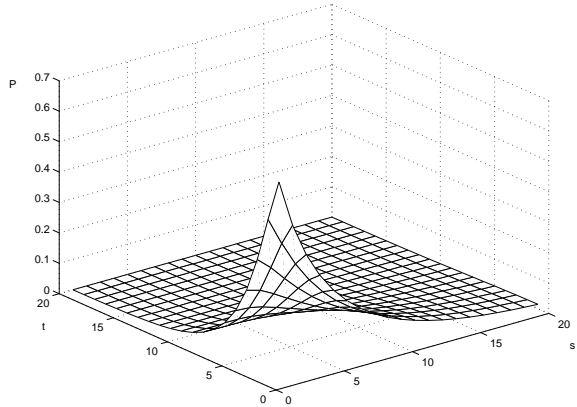}}
\caption[fegf2]{The harmonic oscillator correlation function from Eq (\ref{e16})
as a function of the time $t$ and the variable $s$ for the same 
values of $m=1$ and  $ w_0=0.4$ as in Figure 1. The integral in Eq (\ref{e20}) 
have  been numerically calculated  for values of $t$ and $s$ in the ranges
 $1\le t \le 20$ and $1\le s \le 20$. Note that the correlation tends to zero
 for large $s$ even at those values of $t$ in which the correlation from Eq
 (\ref{e21}) (without $s$) obtains its larger values. }
 \end{figure}

\begin{figure}[hb]
\centerline{
\epsfxsize=5in
\epsffile{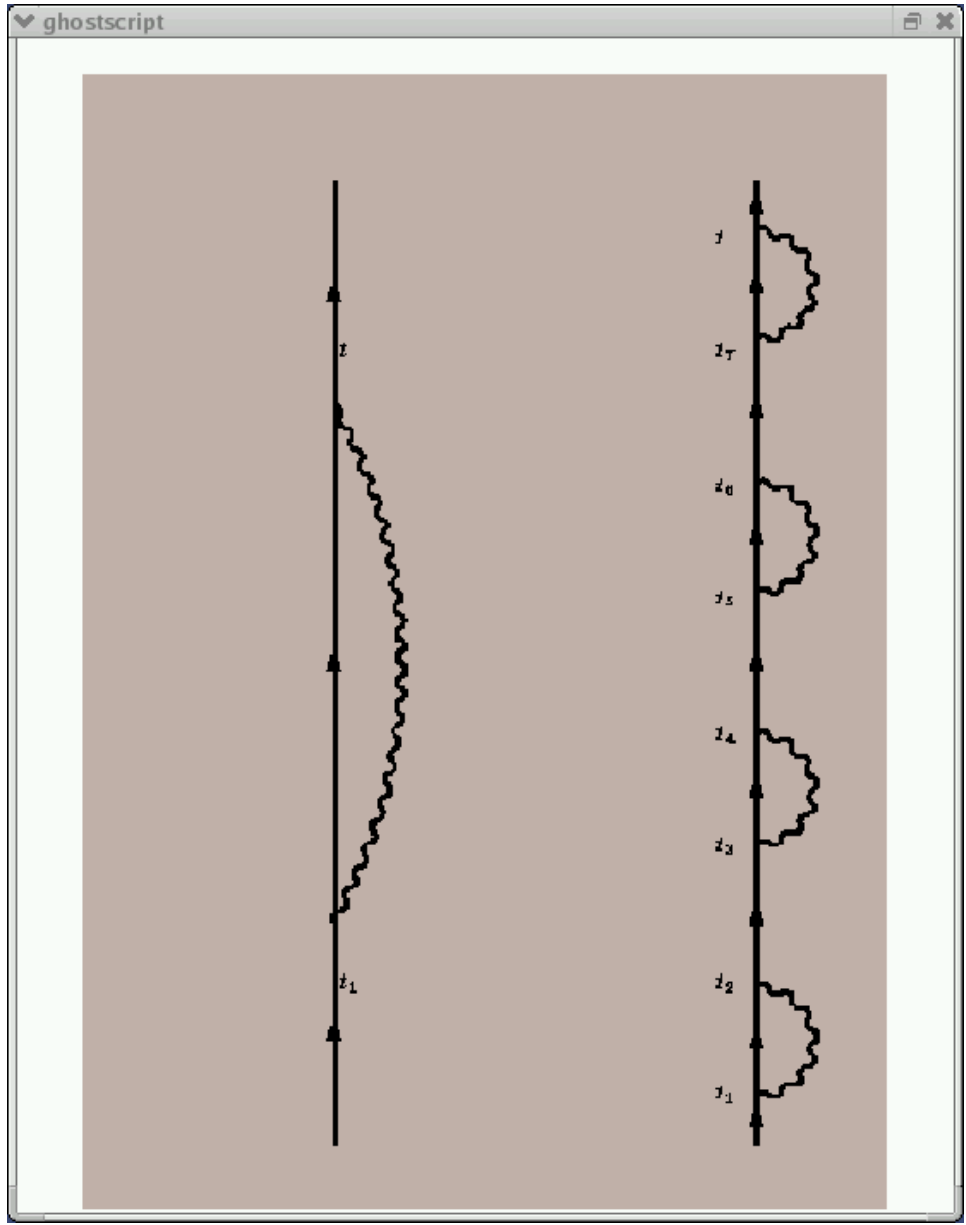}}
\caption[fegf3]{The left hand side of the figure shows the 
 process of emitting and reabsorbing a photon in the time interval
$(t_0,t)$ 
where the energy is not conserved. The electron is represented in the figure by
the directed arrow and the photon by the wavy line.  The right hand side of the
figure shows the same  process repeated four times, in a perturbative
manner,  over the same time interval.  }

\end{figure}

   \end{document}